# Do the $CO_2$ absorption by plants and emission by growing media obey Fick's law?

Nadhifa Zahira Ramadhani Mart[1] and Terry Mart[2‡]

[1]*Departemen Biologi, FMIPA, Universitas Indonesia, Depok 16424, Indonesia*
[2]*Departemen Fisika, FMIPA, Universitas Indonesia, Depok 16424, Indonesia*
[‡]*Email:* terry.mart@sci.ui.ac.id

## ABSTRACT

A simple homemade greenhouse was constructed as a part of the high school project during the learning-from-home period. The greenhouse was used to show $CO_2$ absorption by *Sansevieria trifasciata* during the night, obeying Fick's law, while the $CO_2$ emission by the medium used to grow the plant does not. Although other plants can be used for this purpose, the use of *Sansevieria* is interesting and offers certain advantages because it absorbs $CO_2$ during the night. The project could also provide a simple tool for high school students to investigate the absorption and emission of $CO_2$ from home or in laboratory. Extension of this project to investigate the effect of plant respiration on the environmental parameters, such as decreasing temperature and increasing humidity, is also possible.

## Introduction

The experiment presented in this paper was a project given by *Lazuardi Global Compassionate High School* to the first author during the learning-from-home period due to the COVID-19 pandemic. The high school changed the learning method to a project-based one to overcome the problem of limited time during the online course. The topic of this project was motivated by the trend of gardening hobbies at home during the pandemic followed by the spreading of misleading recommendations for placing *Sansevieria* in the bedroom to alleviate $O_2$ deficiency at night. We know that the latter is not possible. Nevertheless, *Sansevieria* is still appreciated as a $CO_2$ absorber during the night. Although the plant is widely known as *Sansevieria trifasciata*, its official name is *Dracaena trifasciata* (Kew, 2021). For the sake of brevity, in the following we will just use *Sansevieria* when referring to this plant.

*Sansevieria* belongs to the plants that use the crassulacean acid metabolism (CAM) process to absorb $CO_2$ during the night. The CAM process is one of the plant strategies to increase the efficiency in using water, especially in arid environments, such as deserts or habitats with intermittent water supply. In this process the stomata are open during the night and closed during the day. The absorbed $CO_2$ is stored in the form of malic acid in vacuoles and converted back to $CO_2$ during the day for the process of photosynthesis (Cushman, 2001). Therefore, the use of *Sansevieria* gives a special advantage, i.e., we do not need the sunlight during the experiment. As a result, this experiment offers more flexibility. Furthermore, since the greenhouse constructed for this experiment is completely sealed, the use of direct sunlight could tremendously increase the temperature inside the greenhouse and disturb the measurement.

It is also important to note that through this experiment students can learn that the $CO_2$ absorption by plant leaves is a diffusion process and, therefore, follows Fick's law. As a bonus, students can also learn that the medium used to grow the plant could become a strong $CO_2$ emitter. This is the topic of the second experiment.

The experiment was primarily intended for a project at home. However, it can be easily modified for experiment in a school laboratory. This experiment addresses the high school NGSS standards for life and earth sciences, i.e., HS-LS2 Ecosystems: Interactions, Energy, and Dynamics (NGSS Lead States, 2013).

## Fick's First Law

In 1855 Adolph Fick derived two laws describing the diffusion process (Fick, 1855). However, only the first law is related to our present inquiry. To understand this law, let us consider two chambers with different particle concentrations $C_1$ and $C_2$ illustrated in Figure 1. The two chambers are separated by a thin membrane, whose thickness is $\Delta x$. For simplicity, we will assume that the concentrations in both chambers are uniform and inside the membrane the concentration linearly decreases in one-dimensional direction $x$. If $C_1 > C_2$, then there exists a concentration gradient that forces the particle to diffuse from the left chamber to the right one through the membrane. As a result, there is a flux of particle $J$, defined as the number of particles moving through the membrane per unit time and membrane surface area, which is proportional to the concentration gradient. Since the concentration difference can be written as $\Delta C = C(x+\Delta x) - C(x)$, we may write this gradient as $-\Delta C/\Delta x$, where the minus sign indicates that the concentration decreases with increasing $x$ inside the membrane. The proportionality of the gradient to the flux may be written in the form of an equation,

$$J = -D \frac{\Delta C}{\Delta x} \qquad (1)$$

where $D$ is the proportionality constant, which is known as the diffusion coefficient (Nobel, 2009). Equation (1) is the Fick's first law.

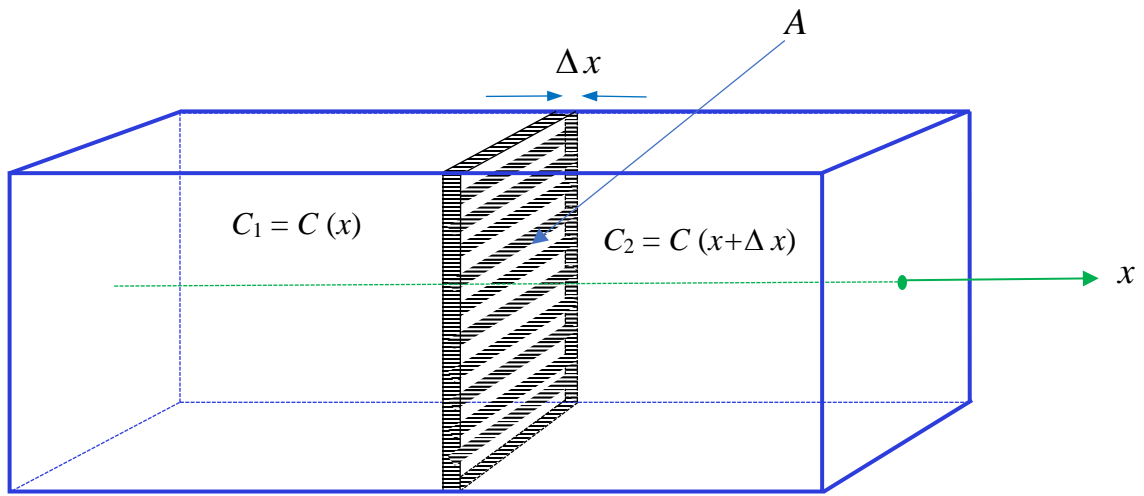

Figure 1. Two chambers containing different particle concentrations $C_1 = C(x)$ and $C_2 = C(x+\Delta x)$, with $C_1 > C_2$, separated by a membrane with thickness $\Delta x$ and surface area $A$.

For the present discussion, it is very convenient to define the particle rate $r$ as the number of particles diffusing through the membrane per unit time, i.e., $r = J A$. Thus, in term of diffusion rate, the Fick's first law reads (Robischon, 2017) (Villani, Dunlop, & Damitz, 2007) (Nobel, 2009) (Evans & von Caemmerer, 1996) (Fick, 1855)

$$r = -A D \frac{\Delta C}{\Delta x} \qquad (2)$$

Equation (2) shows that the rate of diffusion is linearly proportional to the surface area of the membrane and the difference of particle concentrations $\Delta C$. Furthermore, the diffusion rate is inversely proportional to the membrane thickness $\Delta x$. In the present inquiry we assume that both $\Delta C$ and $\Delta x$ are constant. This assumption renders the rate of diffusion given by Equation (2) as merely a linear function of the membrane surface area $A$.

## Tools and Materials Used in the Experiments

In these experiments we need to build a small greenhouse made of cardboard and plastic sheet, and to obtain a small, relatively cheap, $CO_2$ data logger. The homemade greenhouse shown in panel (a) of Figure 2 has dimensions of 0.86 x 0.65 x 0.81 m$^3$, or with a total volume of $V$ = 0.453 m$^3$. The size of the greenhouse is arbitrary, but the minimal size is limited by the dimension of the *Sansevieria* plant used in this experiment. A larger dimension should be avoided, since the measurement of $CO_2$ would be inefficient due to the small absorption rate by the *Sansevieria*. The greenhouse is covered by a plastic sheet to seal the greenhouse from outside air, so that there is no $CO_2$ inflow or outflow during the experiment. Clear plastic tape can be used to tightly seal the greenhouse edges and vertices.

The *Sansevieria* plant, panel (b) of Figure 2, can be bought from a flower store. A bundle of *Sansevieria* plants usually consists of two to five leaves. In this experiment we used 31 leaves, which can be placed in a flowerpot.

We note that there has been an experiment proposed to measure the surface area of leaves by collecting a sample of leaves, measuring their weights, and performing a *t*-test on the sample (Cooper, 2021). Despite its better accuracy, we will not use this experiment since additional experiments would add to the level of difficulty, require more tools (e.g., the metric balance with accuracy of ±0.01 g), and might distract from the focus. Furthermore, we have observed that unlike the leaves of American sweetgum trees (Cooper, 2021), *Sansevieria* leaves have a simpler form. For instance, we sampled the middle part of millimeter graph paper used to measure the Sansevieria leaf area and found that only the left and right sides of the paper were cut, leading to measurement uncertainty, while the top and bottom sides were bounded by the lines of millimeter graph paper. The lengths of the cuts were measured and found to be 100 mm, and the total area of the sample was 2063 mm$^2$. Assuming that we have an uncertainty of 0.5 mm$^2$ for each mm length along the cut, we estimate that the uncertainty obtained by using this direct measurement is 50 mm², which is about 2.5% of the measured area. Thus, we believe that conventional measurement of the surface area of *Sansevieria* leaves by using millimeter graph paper is a reasonable low-cost solution and still sufficiently accurate for the present purpose.

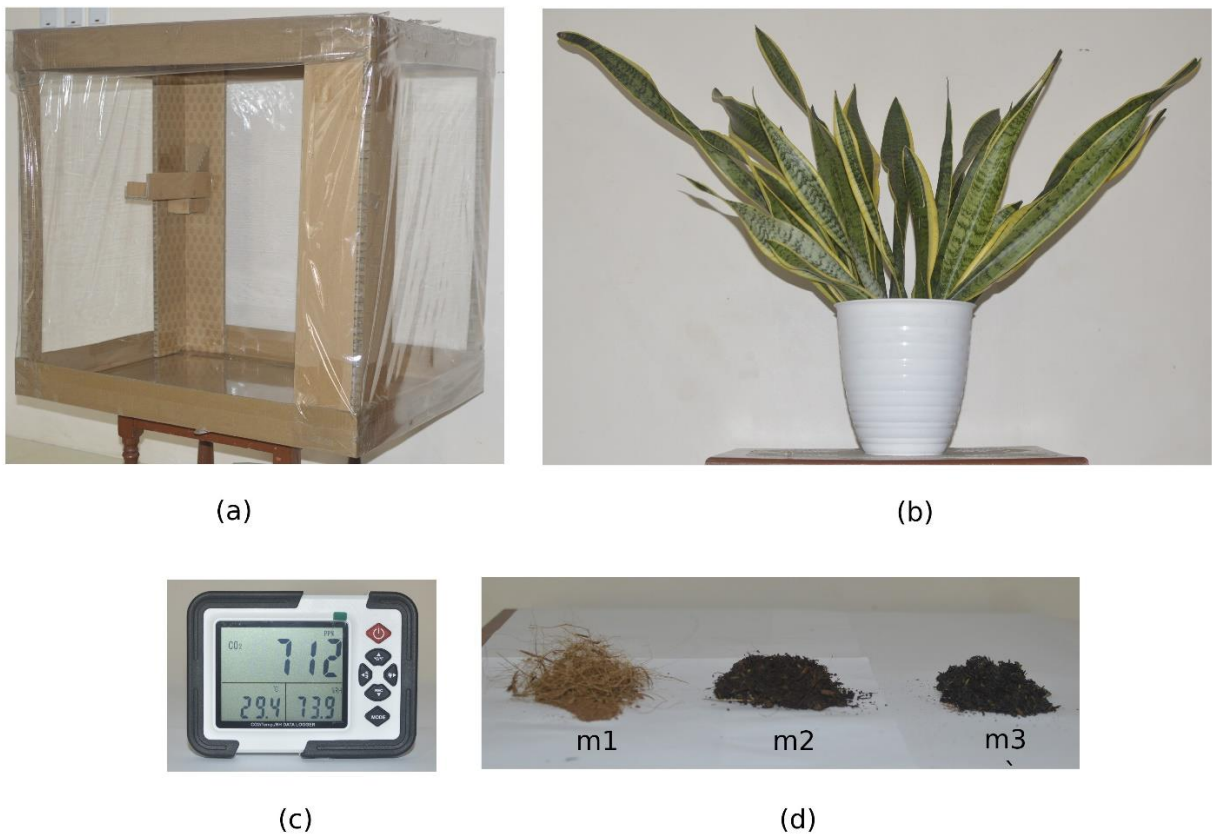

(a)　　　　　　　　　　　　　　　　　　(b)

(c)　　　　　　　　　　　　(d)

Figure 2. Materials and tools used in this experiment: (a) homemade greenhouse, (b) *Sansevieria-trifasciata*, (c) $CO_2$ data logger, and (d) growing media. The growing mediums 1, 2, and 3 are indicated by m1, m2, and m3, respectively. Note that the pictures are not on the same scale.

The $CO_2$ data logger is shown in panel (c) of Figure 2. The brand name is Hti, from Hti Instrument. The instrument model is HT2000 CO2/Temp./RH DATA LOGGER made by Dongguan Xintai Instrument Co., China. This tool is needed to record the $CO_2$ level during the experiment. The range of measurement is 0 - 9999 ppm, with accuracy of ±50 ppm for the range of 0 - 2000 ppm. Besides the level of $CO_2$ this tool also records temperature, humidity, and dew point, which are not used in the present discussion, but could be considered for the extension of this experiment. We note that there are several variants of the data logger available in the market that can be used for this experiment. Ideally, the data logger is provided by the school. Students can borrow it when they are ready to do the measurement. Alternatively, the cost to buy the data logger could be shared among the students, since in principle the measurement will only take five nights per student. Therefore, the cost per student would not be too expensive.

The growing medium used in this experiment can be obtained from a flower shop. In this example we used three different growing mediums, as shown in panel (d) of Figure 2. The material composition of the mediums was not accurately specified on the medium bag. However, in our experiment we noticed that the medium m1 was made of cocopeat and that the medium m3 contained mostly burnt rice husks.

## Experimental Procedures

During the learning-from-home period, all experiments are performed at the student's home. However, they could be also performed at school if there were no COVID-19 pandemic issues.

### Experiment 1. The $CO_2$ Absorption by Sansevieria

In this experiment the $CO_2$ measurement is performed during the night inside the room to avoid extreme temperature changes. There is no specific suggested time to start the experiment. However, since the experiment last for about 8 hours, it is recommended to start the experiment before 9 pm. If the room cannot be made dark, the greenhouse can be covered by black paper sheet or fabric. Note that since it is difficult to fix the initial $CO_2$ level at the same value in each measurement, it is recommended to consider the relative $CO_2$ level, which is obtained from the difference between the actual and the initial $CO_2$ levels. In other words, we set the initial $CO_2$ level to zero.

1. Build the greenhouse frame from cardboard and cover it with a plastic sheet as shown in Figure 2(a). Except its front side, all sides of the greenhouse are completely sealed to avoid the gas inflow and outflow. The front side serves as the window, through which we can place or replace the *Sansevieria* or other required tools. It is also covered by a plastic sheet that can be closed and sealed during the experiment.
2. Prepare 30 – 40 *Sansevieria* leaves. Clean their roots and surfaces by using fresh water before each measurement. Number each leaf and measure its surface area by using millimeter graph paper. Divide the leaves into four groups for four measurements.

3. Place the CO$_2$ data logger inside the greenhouse, and seal the greenhouse. Let the data logger record the CO$_2$ during the night. The suggested sampling time is 30 or 60 seconds. Export the logged data in the CO$_2$ data logger to the Excel format by using a laptop. Copy the Excel data to a special folder made for data analysis. This measurement provides the control data, namely, the fluctuation of the CO$_2$ level inside the greenhouse without *Sansevieria*. Furthermore, this measurement is also important to check whether there is an air leakage in the greenhouse since the CO$_2$ level outside the greenhouse could vary along the night.
4. Clean the interior part of the greenhouse, the data logger, and the flowerpot to hold the *Sansevieria*, to reduce the CO$_2$ contamination from the previous experiment. Use water instead of the growing medium inside the pot to avoid the *Sansevieria* leaves from drying out during the experiment. Place the first group of *Sansevieria* leaves inside the greenhouse. Record the CO$_2$ level inside the greenhouse, and collect the data as in item 3.
5. Repeat the experiment in item 4 by adding another group of *Sansevieria* leaves, until all *Sansevieria* leaves have been used. In total there are one measurement without and 4 experiments with *Sansevieria*. The total time of measurements is five nights.

## *Experiment 2. The CO$_2$ Emission by Growing Media*

This experiment can be performed during the day or night. However, it is recommended to do the experiment inside, where temperature is relatively stable.

1. Prepare three different growing mediums.
2. Prepare four containers for placing the growing medium with different cross-sectional areas. For example, we can use 3, 6, 9, and 12 x 10$^{-2}$ m$^2$.
3. Clean the interior of the greenhouse before performing each measurement. Place 1 kg medium m1 in the container with area 9 x 10$^{-2}$ m$^2$ and the data logger inside the greenhouse. Close and seal the greenhouse. Let the data logger record the CO$_2$ increase for two hours with the sampling time 10 s. Export and copy the obtained data as in Experiment 1.

4. Repeat the measurement in item 3 for media m2 and m3.
5. Indicate the medium that emits the highest $CO_2$ concentration and use it in the following measurements.
6. Repeat the measurement in item 3 for mediums indicated in item 5 with different container cross-sectional areas, i.e., 3, 6, and 12 x $10^{-2}$ $m^2$.

## Results and Discussion

The results shown here provide an example of what students can obtain from this experiment. Students are assumed to be familiar with the software to plot an array of data such as Excel or OpenOffice. The recorded data can be easily retrieved from the $CO_2$ data logger and exported to the Excel format. A step-by-step procedure to record, retrieve, export, and analyze the data, along with the proposed rubric, is provided as Supplementary Material of this paper (available on request). Although the sampling time of data recording has been set to 30 and 10 seconds for Experiments 1 and 2, respectively, it can be increased or decreased to obtain the optimal resolution.

The experimental set up is shown in Figure 3(a). The millimeter graph paper used to measure the surface areas of the Sansevieria leaves are shown in Figure 3(b). The number of Sansevieria leaves used in each measurement is given in the second column of Table 1. By using the information obtained from Figure 3(b) we obtain the third column of Table 1, i.e., the total surface area of Sansevieria leaves in each measurement.

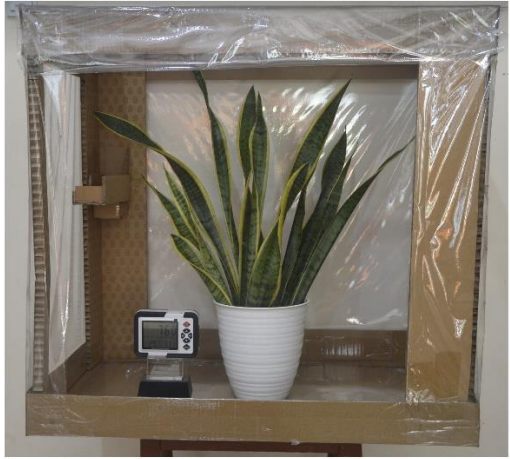
(a)

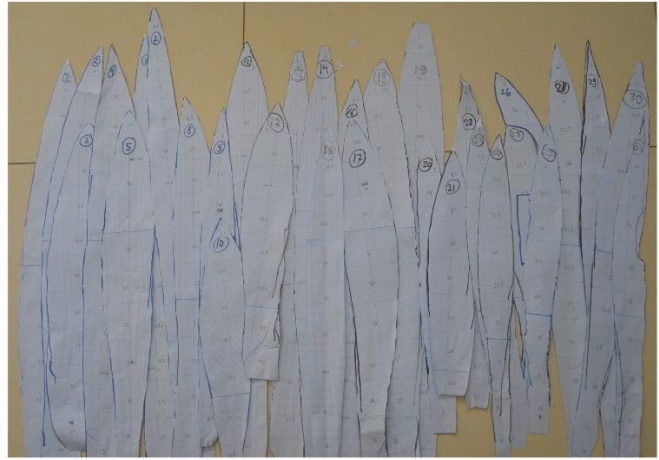
(b)

Figure 3. (a) Experimental setup for measuring $CO_2$ absorption during the night. During the measurement, the greenhouse is covered by a plastic sheet and sealed to avoid gas inflow and outflow, and by papers outside the plastic sheet to minimize the light. (b) To estimate the leaf surface areas, we use the millimeter graph papers.

Table 1. Relation between the number of *Sansevieria* leaves and their total surface area. The average area for each leaf is 0.0243 $m^2$.

| No. | Number of leaves | Total surface area ($m^2$) |
|---|---|---|
| 1 | 8 | 0.2443 |
| 2 | 15 | 0.4066 |
| 3 | 24 | 0.6212 |
| 4 | 31 | 0.7544 |

The result of Experiment 1 is depicted in panel (a) of Figure 4, where the relative $CO_2$ levels inside the sealed greenhouse obtained without and with *Sansevieria* are compared. In the latter, the effect of different numbers of *Sansevieria* leaves on the relative $CO_2$ level is also shown. The experiment without *Sansevieria* provides the proof that the greenhouse is sufficiently sealed. We can clearly see in panel (a) of Figure 4 that in this case the relative level of $CO_2$ is practically unchanged for seven hours of the experiment.

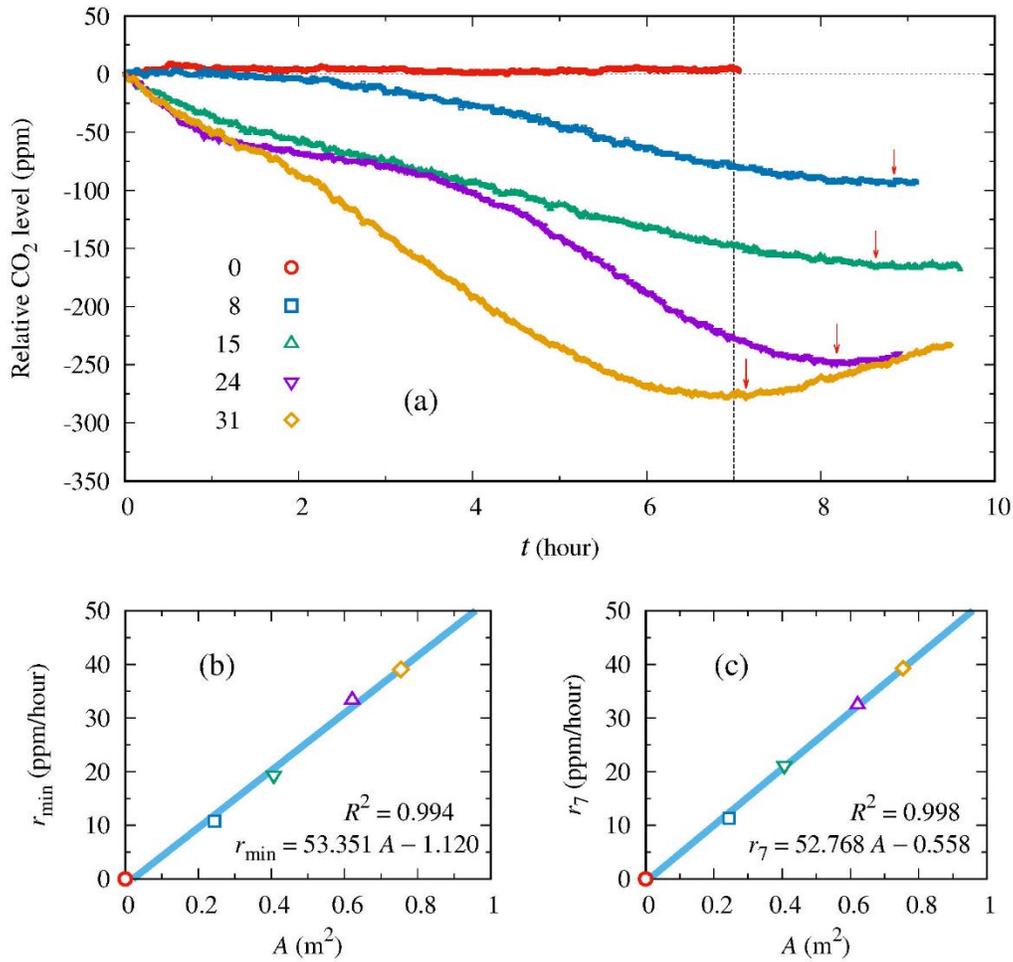

Figure 4. (a) The relative $CO_2$ levels that show the absorbed $CO_2$ by the *Sansevieria-trifasciata* plant as a function of time for different numbers of Sansevieria leaves. The number of leaves is shown in the figure. (b) The average rates of $CO_2$ absorption obtained from panel (a) by using the time required to reach the minimum level indicated by the arrows in panel (a) as a function of the leaves surface area taken from Table 1. (c) Same as in panel (b) but obtained by using $t$ = 7 hours as indicated by the dashed vertical line in panel (a).

The effect of adding *Sansevieria* is clear, the more we place *Sansevieria* leaves inside the greenhouse, the more $CO_2$ gas is absorbed. From panel (a) of Figure 4 it is also seen that the obtained rates do not exhibit a universal pattern. Presumably, it is caused by uneven distribution of $CO_2$ inside the greenhouse during the experiment. In the standard measurement, a small fan can be mounted to help to evenly distribute the air inside the experimental chamber (Nobel, 2009).

Since the absorption patterns are not unique, we have at least two options to determine the *t* value with which we can calculate the rates or the gradients, i.e., (1) using *t* = 7 hours, because we have measured the case without *Sansevieria* for only 7 hours, or (2) using $t_{min}$ at which the relative $CO_2$ level reaches its minimum. The results obtained from the two options are shown in panels (b) and (c) of Figure 4, where it is obvious that they do not produce different results, although the average rates obtained from the 7 hours experiment given in panel (c) show a better linearity. From the linearity relations between the average rates of $CO_2$ absorption and the leaves surface area given in panels (b) and (c) of Figure 4, i.e.,

$$r_{min} = 53.351\,A - 1.120\ , \quad \text{with}\ \ R^2 = 0.994, \qquad (3)$$

and

$$r_7 = 52.768\,A - 0.558\ , \quad \text{with}\ \ R^2 = 0.998, \qquad (4)$$

students can conclude that the absorption of $CO_2$ by the Sansevieria leaves obeys the Fick's law given in Equation (2), since the constants in Equations (3) and (4) are much smaller than the gradients (slopes). Furthermore, they can estimate that the *Sansevieria*, on average, absorbs about 50 ppm per hour per square meter $CO_2$ during the night, or each leaf absorbs $CO_2$ with the rate of approximately 1.7 ppm per hour.

The result of the first measurement of second experiment is shown in Figure 5, where we limit the largest *t* to around 5000 s since our observation shows that the emitted $CO_2$ starts to saturate at this point. Since the differences between the lowest and the largest levels for the three mediums are almost two orders of magnitude, the relative $CO_2$ level must be set into a logarithmic scale, so that all three curves can fit in the same figure. The highest level is shown by medium m3, which mainly contains burnt rice husks.

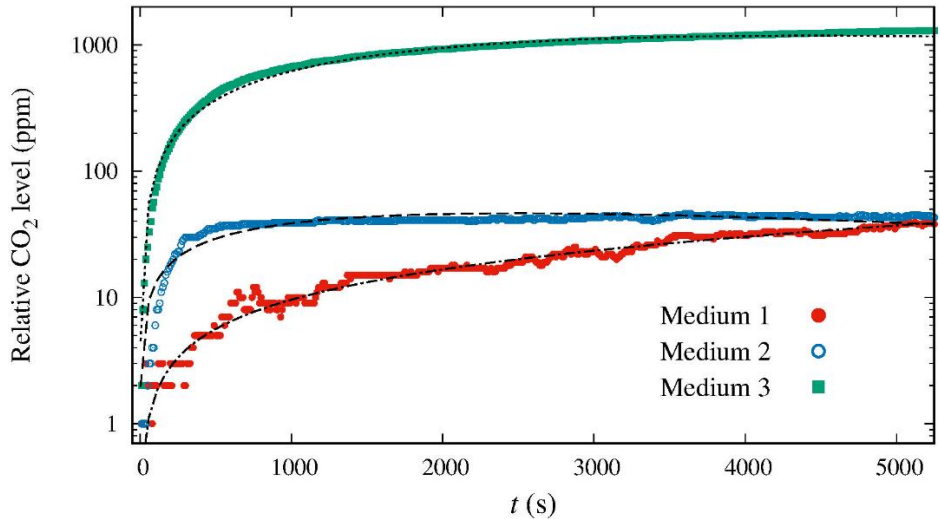

Figure 5. The relative $CO_2$ level detected by data logger that shows the emitted $CO_2$ from three growing media expressed in terms of the relative $CO_2$ level as a function of time. Note that we use logarithmic scale for the vertical axis for the sake of comparison.

For the second measurement the result is shown in Figure 6, in which we have denoted the results obtained from measurement with different areas by A, B, C, and D. In principle, all four curves will eventually converge to the same level due to the same content of the $CO_2$ gas in 1 kg growing medium. However, during the emission process, the emitted $CO_2$ can reach the $CO_2$ data logger with different ways and mechanisms. This could probably explain why in the cases of B and C the $CO_2$ levels do not differ significantly after $t > 1200$ s. Nevertheless, the initial emission rates of the four cases are obviously different and display the dependency on the cross-sectional areas of medium. To investigate this phenomenon, we can focus our attention on the initial rate of emissions, i.e., at $t = 0 - 600$ s. This is shown in panel (b) of Figure 6, where we can see that all curves are almost linearly increasing as a function of $t$. Applying a linear regression to these curves yields the dependency of the rates on the cross-sectional area, which is shown as the gradient $m$ as a function of $A$ in panel (c) of Figure 6. Since $A = 0$ corresponds to no medium, and since the result of the first experiment was that for an empty greenhouse the $CO_2$ level practically does not change, we may assume that $m = 0$ at $A = 0$.

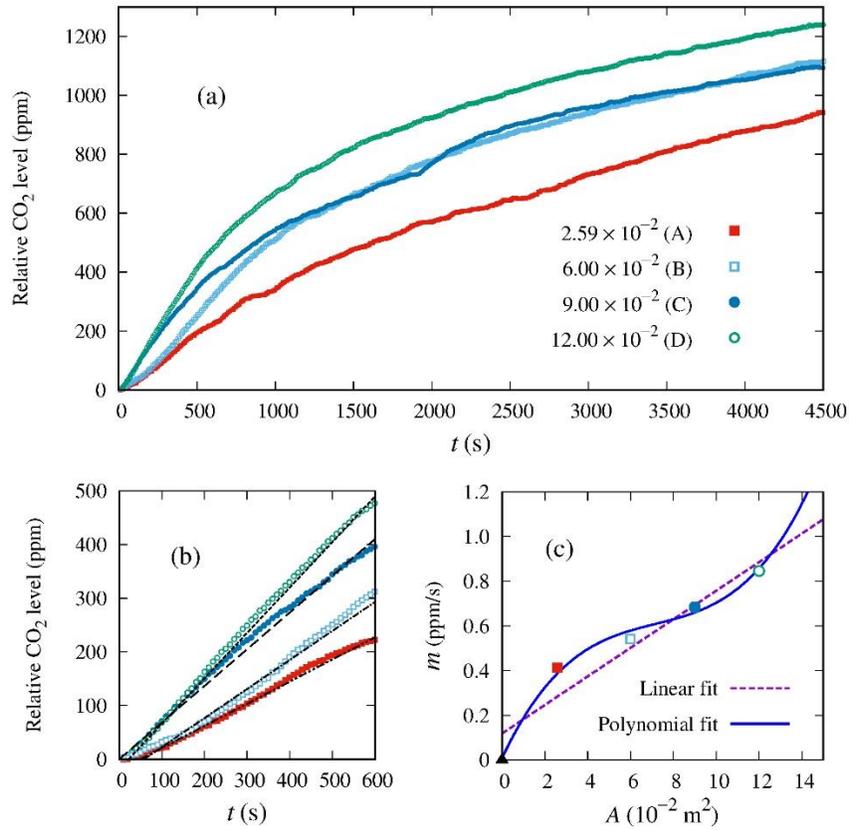

Figure 6. (a) The relative $CO_2$ level detected by data logger that shows the emitted $CO_2$ from the growing medium m3, expressed in terms of the relative $CO_2$ level, as a function of time for four different cross-sectional areas of the medium container. The areas for each curve are shown in the figure and given in units of $m^2$. To facilitate the discussion, we denote the different used areas with A, B, C, and D. (b) As in panel (a) but zoomed for $t$ between 0 and 600 s. (c) Linear and polynomial fits to the gradients $m$ obtained from panel (b) as a function of the container cross-sectional area.

A linear fit to these data yields the dashed line shown in panel (c) of Figure 6, with $R^2$ = 0.918 and a linear equation

$$m = 6.398\, A + 0{,}1181, \qquad (5)$$

which shows that a linear equation does not sufficiently fit the data. A better result would be obtained if we used a third-order polynomial fit in the form of

$$m = 1168.6\, A^3 - 251.28\, A^2 + 20.365\, A + 0{,}0081, \qquad (6)$$

which is also shown in panel (c) of Figure 6 by the solid line. In this case we obtain $R^2 = 0.993$. This result indicates that the emission process in this experiment is not a pure diffusion process, since it is not linearly proportional to $A$. There are at least two possible explanations to this end. First, in the case of the growing medium made from organic waste, the fermentation process yields most of the emitted $CO_2$ gas. Second, in the case of the burnt rice husks, it is possible that there exists $CO_2$ gas trapped in their pores with the internal pressure larger than that of their environment. In the latter, the emission process is driven by the pressure difference.

## Conclusion and Outlook

We have proposed a new method for studying the absorption of $CO_2$ by plant leaves and emission of $CO_2$ from growing media by using a simple homemade greenhouse. It is shown that the absorption of $CO_2$ by plant leaves demonstrates a diffusion process obeying the Fick's law. We used *Sansevieria trifasciata* in this experiment since it absorbs $CO_2$ during the night. By using this homemade greenhouse, we have shown that the growing medium could emit $CO_2$. The emission of $CO_2$ does not follow Fick's law. Thus, we conclude that the emission process is not a diffusion one. The simplicity of the apparatus makes it also possible to use this method in distant learning. Furthermore, by using appropriate sensors, investigation of the absorption or emission of other gasses inside this greenhouse is also possible. The impact of plant respiration on the environmental temperature and humidity could also be studied since the data logger simultaneously records these parameters.

To improve the accuracies of the results, the instructor could ask the student to replicate Step 5 of Experiment 1 with different numbers of Sansevieria leaves and Step 6 of Experiment 2 with different cross-sectional areas. This would be important to confirm the main conclusions of the experiments, namely that the absorption of $CO_2$ by plant leaves shows a diffusion process while the emission of $CO_2$ from growing medium does not. However, it should be remembered that these experiments are performed by each student at home. The first experiment itself would require five nights, provided that all experimental tools worked properly. Therefore, replicating the experiments would be an exhausting task. A possible, but presumably better, solution is to use the experimental results (data) from a group of students who worked separately at their homes. In this case we consider a collaborative analysis by the group of students, which could be performed online during the COVID-19 pandemic.

## Acknowledgment

NZRM would like to thank the teachers of the Lazuardi Global Compassionate High School for their criticisms and suggestions during the experiment and presentation of the result, especially to Mr. Deni Irawan and Ms. Ika Rahayu Sumarni.

## About the Authors

NADHIFA ZAHIRA RAMADHANI MART (nadhifa.zahira21@ui.ac.id) was a student at Lazuardi Global Compassionate High School, Depok 16434, Indonesia. She is now a student in the Department of Biology at the University of Indonesia. TERRY MART (terry.mart@sci.ui.ac.id) is a professor of theoretical physics in the Department of Physics at the University of Indonesia and a member of the Indonesian Academy of Sciences